\documentclass[%
 reprint,
 amsmath,amssymb,
 aps,natbib,
]{revtex4-2}
\usepackage[utf8]{inputenc}
\usepackage{xcolor}
\usepackage{natbib}
\usepackage{appendix}

\usepackage{graphicx}
\usepackage{dcolumn}
\usepackage{bm}
\defcitealias{betranhandy_impact_2020}{Paper~I}
\definecolor{matplotlibblue}{rgb}{0.0, 0.0, 1.0}
\definecolor{matplotlibgreen}{rgb}{0.0, 0.5, 0.0}
\definecolor{matplotlibred}{rgb}{1.0, 0.0, 0.0}
\definecolor{matplotliborange}{HTML}{FF7F0E}
\definecolor{matplotlibmagenta}{rgb}{0.75, 0, 0.75}

\begin{document}
\preprint{APS/123-QED}

\title{A new approximation for heavy-lepton neutrino pair processes in simulations of core-collapse supernovae}

\author{Aurore Betranhandy}
\affiliation{Max Planck Institute for Gravitational Physics (Albert Einstein Institute), Am Mühlenberg 1, D-14476 Potsdam, Germany\\
}%
\author{Evan O'Connor}%
\affiliation{The Oskar Klein Centre, Department of Astronomy, Stockholm University, AlbaNova, SE-106 91 Stockholm, Sweden\\
}%

\date{\today}

\begin{abstract}

In this paper, we present a new approximation for efficiently and effectively including heavy-lepton neutrino pair-production processes in neutrino transport simulations of core-collapse supernovae. In the neutrino-driven explosion mechanism, the electron neutrinos and anti-neutrinos are the main players in transporting the energy of the cooling PNS to the matter behind the shock. While heavy-lepton neutrinos, $\nu_x$, play a smaller role in the heating of the gain region, they dominate the cooling of the proto-neutron star (PNS) and therefore still play a crucial role in the explosion mechanism. In this study, we explore the impacts of modifications in the transport and formalisms of pair ($\nu_x\bar{\nu}_x$) emission and absorption processes. We quantify the impact in terms of the emergent neutrino signal and the nature of the PNS convection and early cooling. For this, we perform a set of simulations, spherically symmetric and axisymmetric, of a 20 M$_\odot$ progenitor using two different state-of-the-art equations of state (EOS). First and foremost, we show that our new efficient approximation for neutrino pair production matches the results of a full treatment very well.  Furthermore, for this progenitor, we show that the impact of the modifications is dependent on the EOS used, as the EOS drives the PNS evolution. The variations we explore, including variations due to the nucleon-nucleon bremsstrahlung formalism, have a comparatively smaller impact than the EOS has as a whole.
\end{abstract}

\maketitle

\section{Introduction}

As one of the ultimate deaths of massive stars, core-collapse supernovae (CCSNe) are important astrophysical phenomena. They also have a large impact on elemental nucleosynthesis \cite{reichert_nucleosynthesis_2021,umeda_nucleosynthesis_2017}. As birthplaces of compact objects such as neutron stars and black holes, they are cornerstones in our understanding of compact objects and compact object mergers. CCSNe are commonly observed through electromagnetic emission but are also sources of neutrinos \cite{abe_first_2022,aartsen_icecube-gen2_2021,abi_supernova_2021}, and due to the inherent multi-dimensionality of these highly energetic events, are also sources of gravitational waves (GWs) \cite{wolfe_gravitational_2023}. To understand current and future observations, theoretical predictions of the different signals emitted by CCSNe are critical. These predictions rely heavily on simulations, as the CCSNe evolution is inherently non-linear. 

Understanding the explosion mechanism behind CCSNe is still an active field of research, even if some theories are becoming more robust, such as the neutrino-driven explosion mechanism or the magnetohydrodynamic (MHD) driven explosion mechanism \cite{burrows_core-collapse_2021}. To explore these mechanisms, we must perform simulations from the onset of the collapse of the star to the explosion itself. As simulations are computationally expensive, they often rely on approximations to reduce the computational cost. Approximations are made in all the different aspects of physics, but here we focus on the ones concerning neutrinos. These approximations are made on the neutrino transport method \cite{mezzacappa_physical_2020}, neutrino interactions with the matter, and neutrino oscillations \cite{liu_universality_2023,balantekin_collective_2024} and can have an impact on the final explosion and the nucleosynthesis \cite{wang_neutrino-driven_2023,wang_nucleosynthetic_2023}. 

As our knowledge of particle physics advances, it is essential to look back at previously accepted approximations and test again their validity and potential impact on our simulations. As an example, while neutrino flavor oscillations in the matter were not included until recently due to their extreme computational cost, new studies have shown that these phenomena can have a significant impact on the final explosion \cite{ehring_fast_2023,nagakura_roles_2023}. Likewise, the constant improvement in the understanding of the interaction of neutrinos with the matter has led to new formalisms for $\beta$-processes through the years \cite{oertel_improved_2020,pascal_proto-neutron_2022,burrows_effects_1998,reddy_neutrino_1998}.

In this paper, we test three different sets of changes for our simulations, part of these changes are extensions of the work done in \citep{betranhandy_impact_2020}, hereafter referred as \citetalias{betranhandy_impact_2020}. First, we explore the impact of different formalisms for the nucleon-nucleon bremsstrahlung pair emission process, one being the One-Pion-Exchange (OPE) formalism from \cite{hannestad_supernova_1998} and the other the T-matrix formalism from \cite{guo_charged-current_2020} in multi-dimensional simulations. Second, we test the calculation of the neutrino-matter interactions for the $\nu_x$ species within our neutrino transport formalism. The main processes for heavy-lepton neutrino creation are the pair processes, namely electron-positron annihilation and nucleon-nucleon bremsstrahlung. The source terms for these interactions can be calculated through a full integration over the neutrino energy, angle, and interaction probabilities or through an effective opacity-emissivity calculation. These two methods are explained and explored in \citetalias{betranhandy_impact_2020}. In this paper, we show that the source term calculation using the effective opacity-emissivity method can be significantly improved by considering the population density of the partner neutrino of the pair. This leads to a third method of transport which we test here. Third, we will test the impact of using two different equations of state (EOS), namely the SFHo EOS from  \cite{steiner_core-collapse_2013} and the baseline SRO EOS \cite{schneider_equation_2019} as used in  \cite{andersen_equation--state_2021}. In the hot CCSN environment, using the SFHo EOS results in a more compact proto-neutron star (PNS) for the same mass \cite{ghosh_pushing_2022,schneider_equation_2020} due to thermal effects. The PNS compactness impacts its cooling, and thus the hydrodynamic quantities at neutrinospheres. These differences arising from the EOS can be used to further quantify the impact of heavy-lepton neutrino cooling. The differences brought by using a different heavy-lepton pair processes formalism will be compared to the impact of using a different EOS.

Our paper is organized as follows, in Sec.~\ref{sec:method}, we will describe the improved approximation for the inclusion of neutrino pair processes in the collision integral. We will also briefly present the numerical codes and EOS used in this paper for simulating the collapse and early post-bounce phase of core-collapse supernovae.
In Sec.~\ref{sec:1D} we will test our improved approximation against the full neutrino pair process implementation in 1D simulations with GR1D. We will also test our implementation in 1D simulations with FLASH \cite{fryxell_flash_2000,andersen_equation--state_2021}. 
In Sec.~\ref{Sec:nu_effect_2d}, we present the results of using the improved approximation and a different formalism for the nucleon-nucleon bremsstrahlung in two-dimensional simulations using FLASH.
We will separate the results into two subsections depending on the EOS, where Sec.~\ref{subsec:SFHo} focuses on simulations using the SFHo EOS, with an early explosion, and Sec.~\ref{subsec:SRO} focuses on simulations using the SRO EOS, with relatively late explosions.
We will then conclude and discuss in Sec.~\ref{sec:Conclusion}.

\section{Method}\label{sec:method}

In this section, we will first present our current approximation used in the neutrino transport for neutrino pair processes, its caveats, and an improvement allowing for a better treatment. We will additionally present different formalisms used for the calculation of the nucleon-nucleon bremsstrahlung interaction from \cite{hannestad_supernova_1998} and \cite{guo_chiral_2019}. We will then present the two simulation codes used in this paper, GR1D and FLASH, and will follow with a presentation of the two different EOS used, the SFHo EOS from \cite{hempel_new_2012,steiner_core-collapse_2013} and the SRO EOS from \cite{schneider_equation_2019,schneider_open-source_2017} which we used to perform simulations of a 20 M$_{\odot}$ progenitor from \cite{woosley_nucleosynthesis_2007}.

\subsection{Approximation of neutrino pair processes in the neutrino transport}\label{subsec:approx}

To compute the neutrino distribution evolution, we need the source terms coming from interactions of neutrinos with the matter. We utilize a standard set of neutrino interactions (see \citetalias{betranhandy_impact_2020} for a complete list).  For this paper we are interested in the transport of heavy-lepton neutrinos and therefore we discuss them in some detail. For these neutrinos, the most relevant interactions are the so-called neutral current interactions.  This includes scattering on nucleons, inelastic scattering on electrons, but also, and most important, pair production processes. This latter process is the main channel for the production and annihilation of heavy-lepton neutrinos. The two neutrino-pair-production processes included in our code are nucleon-nucleon bremsstrahlung (see Section~\ref{sec:brems}) and electron-positron annihilation (see Ref.~\cite{bruenn_stellar_1985}). The difference between these processes arises in their specific interaction probabilities (or kernels). However, the structure of the collision integral, $S^\alpha_{\nu,\mathrm{pair}}$, and the numerical inclusion in the transport, are the same. In this work, the neutrino pair processes are only considered for the $\nu_x$ species of neutrino.

Within the truncated moment formalism, the source term describing the emission and absorption of neutrinos via pair processes is given by \cite{shibata_truncated_2011},
\begin{equation}
S^\alpha_{\nu,\mathrm{pair}}=\nu^3 \int B^\mathrm{pair}_\nu (u^\alpha + l^\alpha ) d\Omega \,,\label{eq:soource_term}
\end{equation}
where
\begin{equation}
    B^\mathrm{pair}_\nu = \int \nu'^2 d\nu' d\Omega' [ (1-f_\nu) (1-\bar{f}_\nu) R^\mathrm{pro} - f_\nu\bar{f}_\nu R^\mathrm{ann}]\,.\label{eq:full_transport}
\end{equation}
In Eq.~\ref{eq:soource_term}, $S^\alpha_{\nu,\mathrm{pair}}$ depends on the energy of the neutrino $\nu$, the four-velocity $u^\alpha$,  a unit vector $l^\alpha$ perpendicular to $u^\alpha$, and the collision term $B^\mathrm{pair}_\nu$.
The full collision term $B^\mathrm{pair}_\nu$ is shown Eq.~\ref{eq:full_transport}, where $\nu'$ represents the energy of the other neutrino in the pair, $f_\nu$ and $\bar{f}_\nu$ the neutrino and other neutrino distributions respectively, and $R^\mathrm{pro/ann}$ represents the interaction kernel for production and annihilation, respectively and depends on the energy of both neutrinos as well as the angle between them. The use of the full source term derived from Eqs.~\ref{eq:soource_term} and \ref{eq:full_transport} in terms of the radiation moments (see Appendix \ref{sec:appendix}) in our simulation is computationally expensive. When used in this paper, we refer to it as the `Full' method.  Due to the computational expense, simplified and less expensive versions of the collision integral have been created. A useful approximation is to model these interactions as an effective absorption/emission process using a source term as follows,
\begin{equation}
S_{\nu,\mathrm{pair}}^{\alpha} = (\eta_{\nu} - \kappa_{\nu,a} J_\nu) u^\alpha - \kappa_{\nu,a} H_\nu^\alpha\,.
    \label{eq:simplified}
\end{equation}
In this expression, $J_\nu$ and $H_\nu^\alpha$ are the evolved zeroth and first radiation moments, the energy density and momentum density, respectively. $\eta_\nu$ is an emissivity and $\kappa_{\nu,a}$ is an absorption opacity.  As we show in Appendix \ref{sec:appendix}, we can derive effective emission and absorption coefficients for neutrino pair-production processes using a few reasonable assumptions. Assuming no final-state blocking of the emitted neutrinos allows $\eta_\nu$ to be calculated a priori and tabulated (see Eq.~\ref{eq:emissivity_moment}). In principle, $\kappa_{\nu,a}$ depends on the details of the partner neutrino distribution, (see Eq.~\ref{eq:kappanua}). In this work, we take two approaches to calculating this effective opacity.\\

Method \#1: Detailed balance ensures that if the radiation fields are in equilibrium with the matter, then any emission must be balanced by an equal absorption.  Therefore, our first approximation assumes $\kappa_{\nu,a} = \eta_\nu / J_{\nu, \mathrm{eq}}$, where $J_{\nu, \mathrm{eq}}$ is the equilibrium distribution, this forces Eq.~\ref{eq:simplified} to zero in this regime. This is akin to Kirchhoff's law for single neutrino processes. Like the emissivity, this opacity can be calculated a priori and tabulated. This method has been used and tested in \cite{oconnor_open-source_2015,betranhandy_impact_2020}, we refer to this as our `Reference' method.\\

While the reference method above gives a $\kappa_{\nu,a}$ that is correct when the neutrino fields are the equilibrium fields, this is not true when the fields are out of equilibrium (unlike for Kirchhoff's law for single neutrino interactions where the absorption opacity is valid regardless of the neutrino distribution).  This is because the reference method above is equivalent to assuming an equilibrium distribution for the partner neutrino when calculating the absorption opacity (see Eq.~\ref{eq:kappanua}). 

While this approximation is valid in the core of the PNS, the assumption of the equilibrium field for the other neutrino breaks when the neutrino begins to decouple.  
\begin{figure}
    \centering
    \includegraphics[width=\linewidth]{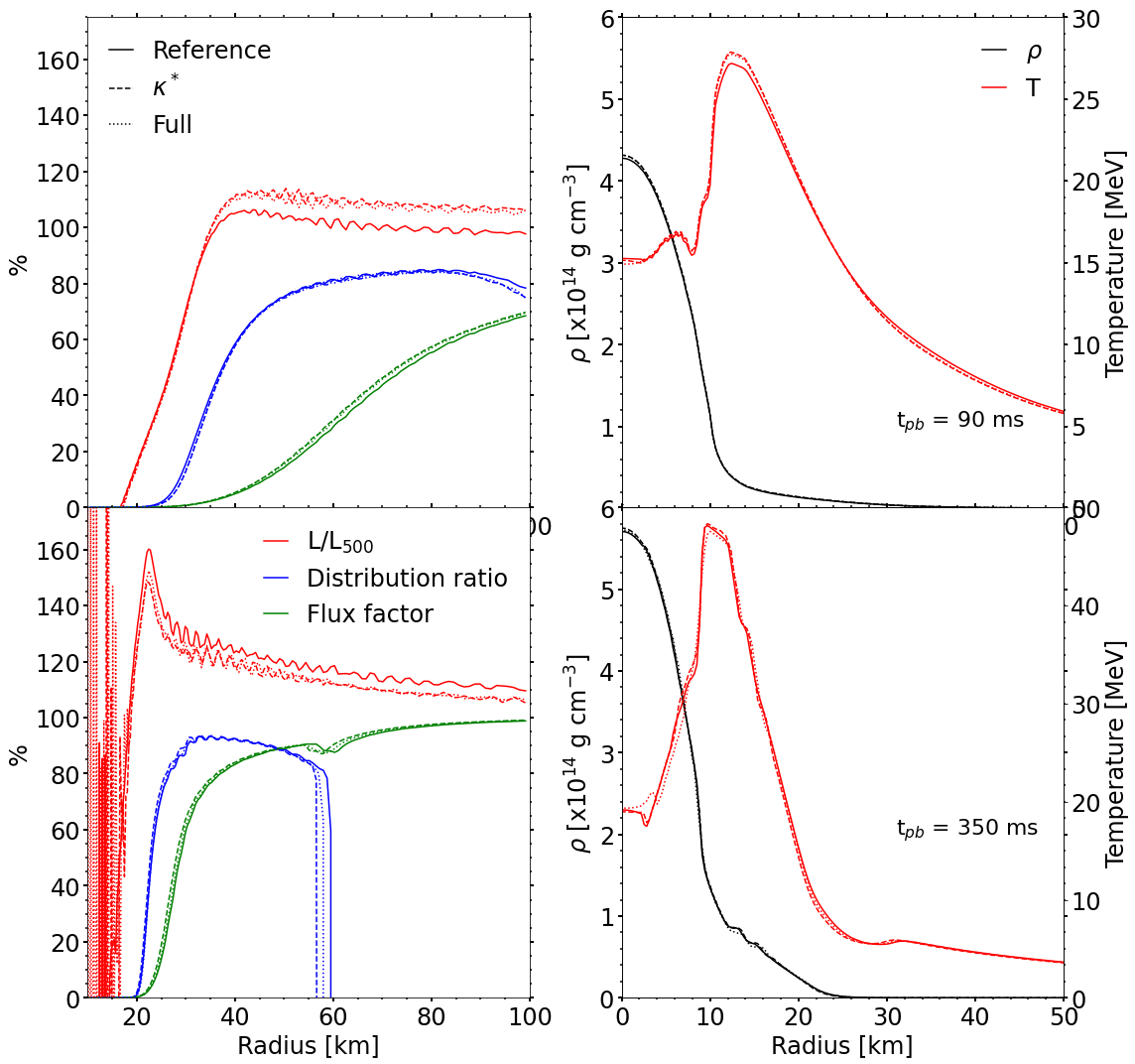}
    \caption{Radial neutrino and hydrodynamic properties at early (top) and late (bottom) times for the 20\,M$_\odot$ progenitor. We show results for the kernel treatment of pair processes with dotted lines, the simplified pair processes treatment with solid lines, and our modified absorption opacity $\kappa_a^*$ with dashed lines. In red, we show the radial evolution of the total outgoing neutrino luminosity, normalized to the value of the kernel treatment at 500\,km. In blue, we show the relative difference between the actual and equilibrium neutrino distribution function, while in green we show the flux factor for neutrinos in the energy bin centered at 15 MeV. The hydrodynamic properties are shown in the right panel of the figure, namely the radial evolution of the density and the temperature for the different simulations at early and late times.}
    
    \label{fig:lumi_factor}
\end{figure}
We show this in Fig.~\ref{fig:lumi_factor} (modified Fig.~2 of \citetalias{betranhandy_impact_2020}), where we see the deviation in the radial profiles of the luminosities (left panels; red lines) between simulations using the full method (Full, dotted) and simulations using method \#1 (Reference, solid lines) for the computation of neutrino pair processes begins where the neutrino distribution deviation from the equilibrium distribution (blue lines; $1-J_\nu/J_{\nu, \mathrm{eq}}$) is $\sim$50\%. Here, method \#1 overestimates the absorption opacity because the radiation field of the partner neutrino is no longer the equilibrium field, but rather only a fraction of it.  This has the effect of overestimating the opacity $\kappa_{\nu, a}$ and leads to an excess absorption and a lower emergent neutrino luminosity. The deviation in luminosity has a direct impact on the temperature at the surface of the PNS in the early cooling phase, as seen in the right panel of Fig.~\ref{fig:lumi_factor}. This can influence the production of electron neutrinos and anti-neutrinos and lead to changes in the neutrino heating. \\

Method \#2: To improve the reference version (method \#1 above) of the neutrino pair process calculation while not sacrificing computational cost by using the full collision integral, we modified the implementation of the heavy-lepton neutrino absorption opacity by a multiplication factor designed to reduce the absorption when the partner neutrino field is not equal to the equilibrium distribution. This simple, energy-independent, factor is defined as follows, 
\begin{align}
    \kappa_{\nu,a}^* &= F \kappa_{\nu, a}, \label{kappa_mod}\\
    F &= \mathrm{min}\left(1,\frac{\sum_{i} \overline{E_i} \Delta \epsilon_i}{\sum_{i} \overline{J}_{i,\mathrm{eq}} \Delta \epsilon_i}\right). \label{simp_mod_factor} 
\end{align}
\noindent with $\kappa_{\nu,a}$ the neutrino opacity determined by method \#1 above, $\overline{E}_i$ the spectral neutrino energy density of the partner neutrino and $\overline{J}_{i, \mathrm{eq}}$ the neutrino equilibrium spectral energy density of the partner neutrino. The minimum factor shown in Eq.~\ref{simp_mod_factor} is for numerical stability but is rarely used as the neutrino fields should never exceed the equilibrium distribution. Therefore, this factor, $F$, is always equal to or smaller than 1, effectively reducing the opacity. Integrating over all the energy bins is necessary to reproduce the ability of a neutrino of specific energy to interact with anti-neutrinos of all other energies. When the neutrino energy density is in equilibrium, $F=1$, we recover detailed balance. This is an empirical improvement to our effective opacity and one that, as we shall see below, does a significantly better job capturing not only the emergent neutrino spectrum but also the radial profiles of the neutrino luminosity and the resulting hydrodynamic properties.  We refer to this method throughout the paper as the `$\kappa^*_{a}$' method. For simplicity, from here on we omit the $\nu$ subscript, although these emissivities and opacities are still energy dependent. 

\subsection{Bremsstrahlung formalism}
\label{sec:brems}
In this paper, we use two different formalisms for the nucleon-nucleon bremsstrahlung neutrino pair process. The formalisms are the same as the ones used in \citetalias{betranhandy_impact_2020}, e.g, the One-Pion-Exchange (OPE) formalism from \cite{hannestad_supernova_1998}, and the T-matrix formalism from \cite{guo_chiral_2019}. We follow the procedure outlined in Section \ref{subsec:approx} for determining $\kappa^*_a$ and $\kappa_a$ for use in our approximate scheme.

\subsection{Setup}

To verify the implementation of our modified absorption opacity $\kappa^*_a$, we must compare it to the full transport results for the heavy-lepton neutrino luminosity and mean energy. This leads us to test our implementation in GR1D, where all the different transport methods are implemented, and then use it in FLASH to perform multi-dimensional simulations. 

\subsubsection{GR1D}\

GR1D is a fully general relativistic (GR) hydrodynamic code in spherical symmetry \citep{oconnor_new_2010,oconnor_open-source_2015}. 
In GR1D, we use the same parameters as the ones used in \citetalias{betranhandy_impact_2020}. The size of the simulation is fixed at 800 zones logarithmically spaced, with the maximum radius being fixed at the radius for which $\rho$ = 10$^2$ g\,cm$^{-3}$. The inner 20\,km has a constant grid spacing of 300~m. GR1D uses an M1 scheme with an analytic closure from \cite{cardall_conservative_2013} for the neutrino transport. The neutrino energy dimension has been discretized into 18 bins.  

\subsubsection{FLASH}\

FLASH is a hydrodynamic simulation code \citep{fryxell_flash_2000,couch_revival_2013,oconnor_exploring_2018} using a modified GR effective potential in Newtonian hydrodynamics \citep{marek_exploring_2006}. While some GR effects are included in the neutrino transport, this is not the case in the hydrodynamics other than the use of the effective potential. 
FLASH uses a uniform grid combined with an adaptive mesh refinement scheme. For our 1D simulations, the maximum radius has been fixed to 10$^9$\,cm and the maximum refinement level to 10. This gives a resolution of $\sim$203\,m in the center of the PNS. The angular resolution is fixed at 0.5°.\\
For our 2D simulations, the maximum radius is fixed at 10$^9$\,cm in the x-direction while it spans from -10$^9$\,cm to 10$^9$\,cm in the y-direction. For simulations using the SFHo EOS, the minimal size of a cell is 407\,m in the center of the PNS. For simulations using the SFHo EOS, the minimal size of a cell is 465\,m in the center of the PNS. We have the same maximum angular resolution of 0.5° as for the 1D simulations. \\
 FLASH also uses an M1 scheme for neutrino transport, using 18 energy bins for the 1D simulations and 12 energy bins for the 2D simulations. The neutrino physics is the same as the one used in GR1D, with a pre-computed opacity table generated with NuLib. 

\subsubsection{EOS}\

We use two different EOS in this study. First, the SFHo EOS from \cite{steiner_core-collapse_2013}, to compare the results to \citetalias{betranhandy_impact_2020}, and second, the SRO EOS from \cite{schneider_equation_2019} to test the EOS dependence.  The SFHo EOS, having a more compact core for the same mass \cite{ghosh_pushing_2022,schneider_equation_2020}, due in part to the thermal part of the EOS, impacts the cooling of the PNS and the hydrodynamic quantities at neutrinospheres. These two EOS can be used to further test the impact of heavy-lepton neutrino cooling versus the impact of using a different EOS.

\section{Results}

In this section, we present the results of our simulations using different transport methods and nucleon-nucleon bremsstrahlung interaction formalisms. We will first begin by presenting the results of our 1D simulations, using GR1D and FLASH, to test the implementation and effectiveness of our modified absorption opacity $\kappa_a^*$. We will then present the results of our 2D simulations using FLASH. As the explosion time varies greatly between the simulations using SFHo and the one using the SRO EOS, we choose to present these results separately.

\begin{table}
    \centering
    \begin{tabular}{|c|c|c|c|c|c|c|} \hline  
           &\multicolumn{3}{|c|}{SFHo}&  \multicolumn{3}{|c|}{SRO}\\ \hline  
 & Reference& $\kappa^*_a$ &Full& Reference&$\kappa^*_a$ &Full\\ \hline  
           OPE&{\color{matplotlibred} \huge\textbf{-}}&  {\color{matplotlibred} \huge\textbf{-}\hspace{1mm}\huge\textbf{-}}&\huge\textbf{\color{matplotlibred} $\cdot$  $\cdot$}&  {\color{matplotlibblue} \huge\textbf{-}}& \huge\textbf{\color{matplotlibblue} -\hspace{1mm}-}&\huge\textbf{\color{matplotlibblue}  $\cdot$ $\cdot$}\\ \hline  
           T-matrix&\huge\textbf{\color{matplotliborange} -}&  \huge\textbf{\color{matplotliborange} -\hspace{1mm}-}&\huge\textbf{\color{matplotliborange}  $\cdot$ $\cdot$}&  \huge\textbf{\color{matplotlibgreen} -}& \huge\textbf{\color{matplotlibgreen} -\hspace{1mm}-}&\huge\textbf{\color{matplotlibgreen}  $\cdot$ $\cdot$}\\ \hline 
    \end{tabular}
\caption{Line style and colors for the different simulations in this work. Simulations using the SRO EOS are represented by green and blue lines, while simulations using the SFHo EOS are represented by red and orange lines. Simulations using the OPE formalism for the nucleon-nucleon bremsstrahlung are represented by blue and red lines, while the ones using the T-matrix formalism are represented by green and orange lines. The linestyle shows the type of transport used for pair processes, with solid line representing the reference transport, dashed lines representing the improved $\kappa^*_a$ transport method, and dotted lines representing the full transportation for pair processes. }

\label{tab:linestyle-color}
\end{table}

\subsection{1D}\label{sec:1D}

\begin{figure}
    \centering
    \includegraphics[width=1.\linewidth]{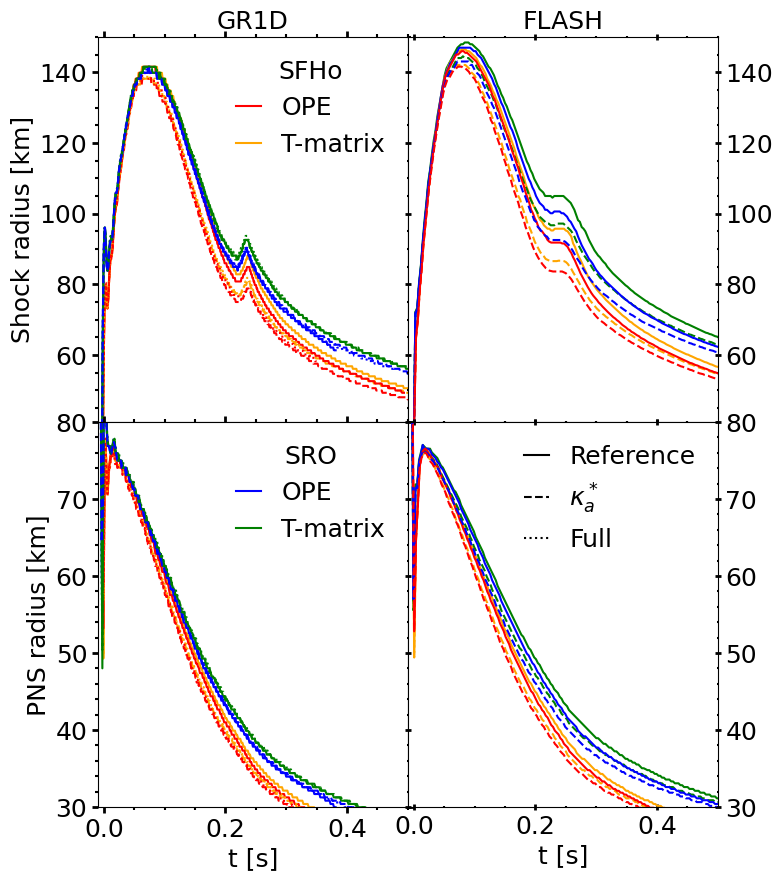}
    \caption{Hydrodynamic quantities for the different 1D simulations. The top row shows the shock radius evolution, and the bottom represents the PNS radius evolution. The left column shows results for simulations using GR1D, while the right column represents the simulations performed using FLASH. Different colors correspond to different EOS and nucleon-nucleon bremsstrahlung formalisms. The blue and green curves are simulations using the SRO EOS, while the orange and red curves are simulations using the SFHo EOS. The red and blue curves are simulations using the OPE formalism for the nucleon-nucleon bremsstrahlung process. In contrast, the orange and green curves use the T-matrix formalism for this interaction. Different line styles correspond to different transport methods. Solid lines indicate the reference transport for the neutrino pair processes, the dashed line represents the simulations using the modified absorption opacity $\kappa^*_a$ for the transport, and the dotted lines represent simulations using the full transport method for pair processes. Simulations using FLASH do not have the latter, as this transport method is not included within FLASH. All these colors and line styles are summarized in Tab.~\ref{tab:linestyle-color} }
    \label{fig:1d_full_hydro}
\end{figure}

We use GR1D to compute simulations showing the difference between a kernel-based treatment of $\nu_x$ pair production (labeled Full), an effective opacity-based treatment (labeled Reference), and our improved effective opacity-based treatment with $\kappa_a^*$ (labeled  $\kappa_a^*$). We also implement this new algorithm into FLASH. In these spherically symmetric simulations, we do not use any artificial explosion mechanism. In Fig.~\ref{fig:1d_full_hydro}, we show the shock radius evolution (top panels) and PNS radius evolution (bottom panel). On the left, we can see simulations performed using GR1D; on the right, simulations performed using FLASH. The red and orange lines represent simulations performed using the SFHo EOS, and the blue and green lines represent simulations performed using the SRO EOS. Dotted lines represent simulations using the full transport method for the neutrino pair processes presented Eq.~\ref{eq:full_transport}, solid lines represent the effective opacity-based transport shown Eq.~\ref{eq:simplified}, and dashed lines represent simulations using our improved effective opacity $\kappa^*_a$ presented Eq.~\ref{simp_mod_factor}. Tab.~\ref{tab:linestyle-color} summarizes the different line styles and colors used throughout the paper.

Starting with the shock radius evolution for simulations using GR1D (top left panel), we observe a difference starting at around $\sim$50\,ms post-bounce when the peak shock radius occurs.\\
For simulations using the SRO EOS, the shock reaches a maximal radius of $\sim$142\,km, showing little to no differences based on the transport method or the nucleon-nucleon bremsstrahlung formalism.  The largest difference arises from using the T-matrix formalism for nucleon-nucleon bremsstrahlung, as seen in \citetalias{betranhandy_impact_2020}. Simulations using the T-matrix formalism always have a slightly larger shock radius. 
For simulations using the SFHo EOS and the reference transport method, the peak radius is the same as the SRO EOS at $\sim$142\,km, independent of the nucleon-nucleon bremsstrahlung formalism. Simulations using the modified absorption opacity $\kappa^*_a$ and the full transport method with the SFHo EOS show a peak shock radius of $\sim$139\,km. This represents a difference of a few percent, which is not so relevant at peak, but this difference increases with time, up to $\sim$10\% at the end of our simulations. The nucleon-nucleon bremsstrahlung formalism shows little impact, with a trend for the simulations using the T-matrix formalism having a slightly larger shock radius than their equivalents using the OPE formalism, the same trend observed with the SRO EOS.\\

In the top right corner of Fig.~\ref{fig:1d_full_hydro}, we show the shock radius evolution for simulations using FLASH. As the full transport method for neutrino pair processes is not implemented in FLASH, only the simulations using the modified absorption opacity $\kappa^*_a$ and the reference transport method are shown. The effects of the different transports and nucleon-nucleon bremsstrahlung formalisms are similar and consistent with the ones observed in simulations using GR1D. The shock radius in simulations using FLASH has a maximal radius slightly higher than in simulations using GR1D. This can be linked to the different gravity treatment, GR for GR1D \cite{oconnor_new_2010} and a modified Newtonian potential for FLASH \cite{fryxell_flash_2000,marek_exploring_2006,oconnor_exploring_2018}. For simulations using the SRO EOS, the shock radius is larger than for simulations using the SFHo EOS, up to $\sim$11\% when comparing the same neutrino treatments. The differences linked to the transport method are comparable for both EOS, with the simulations using the reference transport method having a larger shock radius by up to $\sim$10\%  compared to simulations using the modified opacity $\kappa^*_a$. The differences linked to the nucleon-nucleon bremsstrahlung formalism are lower, with simulations using the T-matrix formalism having a shock radius larger than simulations using the OPE formalism by a maximum of $\sim$5\%.\\

With FLASH, the differences arising from the EOS, neutrino pair processes transport method, and bremsstrahlung formalism are generally larger than the ones observed using GR1D. This could be linked to the different gravitational potential, leading to different PNS evolution. As observed in \citetalias{betranhandy_impact_2020} and \cite{betranhandy_neutrino_2022-1}, a difference in the PNS evolution can impact the shock evolution, especially in one-dimensional simulations.

The bottom row of Fig.~\ref{fig:1d_full_hydro} shows the PNS radius evolution for simulations using GR1D (left) and FLASH (right). All the simulations show a similar PNS radius at peak, but the contraction of the PNS radius for simulations using FLASH is slower than for simulations using GR1D. Simulations using GR1D reach a PNS radius of $\sim$30\,km at $\sim$300-400ms post-bounce, but simulations using FLASH reach the same value at $\sim$400-500ms post-bounce. Depending on the EOS, the PNS radius evolution shows a similar trend independently of the code. The simulations using the SRO EOS systematically reach a 30\,km radius $\sim$100\,ms after the ones using the SFHo EOS. The differences in the PNS radius evolution linked to the transport method are smaller, but simulations using the reference transport method have a slightly larger PNS radius at equal time. The differences linked to the nucleon-nucleon bremsstrahlung formalism are negligible, especially in simulations using GR1D, with a trend of simulations using the T-matrix formalism always having a larger PNS radius. 

\begin{figure}
    \centering
    \includegraphics[width=1\linewidth]{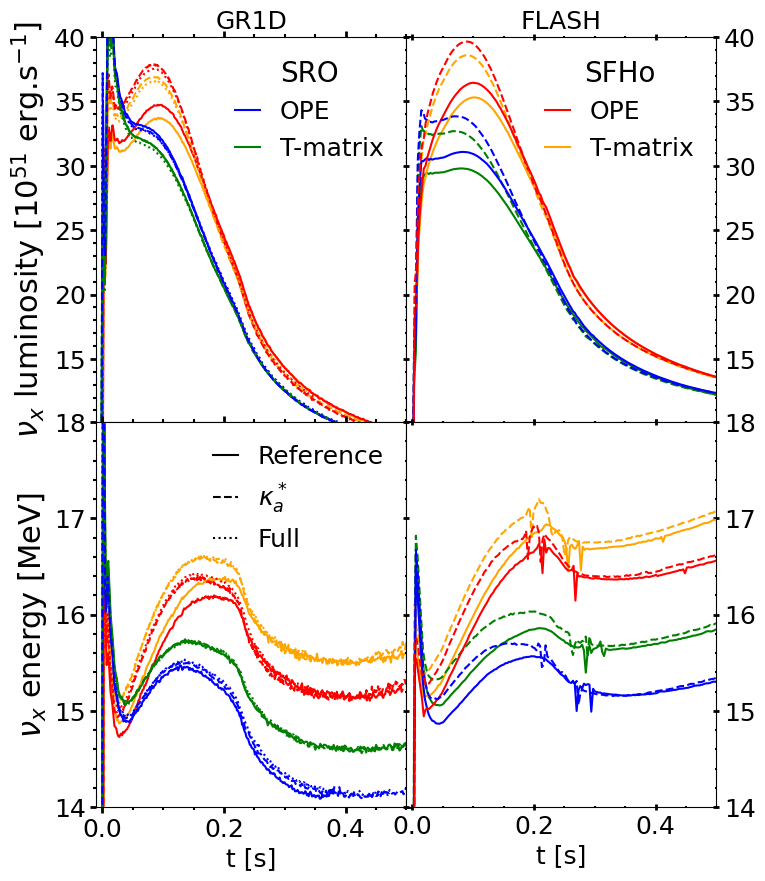}
    \caption{Heavy-lepton neutrino quantities for the different 1D simulations. The top row shows the $\nu_x$ luminosities, while the bottom row shows the $\nu_x$ mean energy. The placement, color code, and line style are otherwise the same as the ones used in Fig.~\ref{fig:1d_full_hydro}. The relatively small glitches in the mean energy for the FLASH simulations arise when the shock crosses a mesh refinement boundary and are due to the velocity dependent effects in the transport.}
    \label{fig:1d_full_nu}
\end{figure}

All these variations can be explained using Fig.~\ref{fig:1d_full_nu}, in which we show the heavy-lepton ($\nu_x$) neutrino luminosities (top row) and mean energies (bottom row).  The placement and color code is the same as the one used in Fig.~\ref{fig:1d_full_hydro}.  We begin our discussion with the $\nu_x$ luminosity in simulations using GR1D (top left panel). The $\nu_x$ luminosities show the largest differences when varying the EOS. 
Simulations using the SRO EOS have a higher neutrino luminosity by $\sim$10-20\% at neutrino burst compared to simulations using the SFHo EOS. This can be linked to the PNS radius, which is larger for simulations using the SRO EOS at this time. The neutrino transport method in simulations using the SRO EOS shows little impact ($\lesssim$1 \%), but the nucleon-nucleon bremsstrahlung formalism causes changes in the $\nu_x$ luminosity between $\sim$30\,ms post-bounce and $\sim$170\,ms post-bounce. During this time, simulations using the OPE formalism have a higher luminosity, up to $\sim$5\%, than simulations using the T-matrix formalism. For simulations using the SFHo EOS, differences are larger, with simulations using the reference transport method having a lower luminosity by $\sim$10\% when compared to simulations using the modified absorption opacity $\kappa^*_a$ or the full transport method. These differences stay through the first $\sim$170ms post-bounce. Simulations using the full transport method and those using the modified absorption opacity $\kappa^*_a$ show similar results. This is a key result and supports the use of our modified effective opacity in place of the full treatment. Simulations using the T-matrix formalism for nucleon-nucleon bremsstrahlung have a lower luminosity by $\sim$5\% compared to those using the OPE formalism for this interaction. Generally, higher $\nu_x$ luminosities give more PNS cooling and therefore lower shock radii (see Fig.~\ref{fig:1d_full_hydro}).
For simulations using FLASH, the $\nu_x$ luminosity (top right panel) shows a higher luminosity at peak by $\sim$10\% for simulations using the modified absorption opacity $\kappa^*_a$ independently of the EOS used. Similarly, using an OPE formalism raises the luminosity by $\sim$5\% at peak. The differences arising from using a different EOS are stronger when using FLASH than when using GR1D. Simulations using the SRO EOS have a peak luminosity lower by $\sim$20\% when compared to the peak luminosity of simulations using the SFHo EOS. Once again, a higher $\nu_x$ luminosity gives a lower shock radii.\\

On the bottom row of Fig.~\ref{fig:1d_full_nu}, we show the $\nu_x$ mean energies for simulations using GR1D (left) and FLASH (right). Beginning with GR1D, we note first that all variations caused differences of at most 2 \,MeV in the mean energies. We separate the impact of the EOS, bremsstrahlung formalism, and transport method. Simulations using the SFHo EOS have a higher $\nu_x$ mean energy by $\sim$7\% compared to simulations using the SRO EOS. The $\nu_x$ mean energy peak is later for the simulations using the SFHo EOS, with a peak at $\sim$150ms post-bounce, while simulations using the SRO EOS show a $\nu_x$ mean energy peak at $\sim$130ms post-bounce. The impact of a different formalism for the nucleon-nucleon bremsstrahlung is similar for both EOS, with simulations using the T-matrix formalism having a higher mean energy by $\sim$3\% throughout the simulations. The transport method produces larger differences for simulations using the SFHo EOS, with a $\nu_x$  mean energy $\sim$2\% higher than the reference method when using the full transport or the modified opacity $\kappa^*_a$. This difference is only present for the first $\sim$250ms post-bounce. When using GR1D, the $\nu_x$ mean energy for simulations using the modified absorption opacity $\kappa^*_a$ is close or identical to the $\nu_x$ mean energy of simulations using the full transport method. Simulations using the SRO EOS show little impact from the use of different formalisms for the nucleon-nucleon bremsstrahlung or the use of different transport methods for the pair processes. When using FLASH, the differences linked to using different EOS and formalism are similar to the ones appearing when using GR1D. The differences linked to the transport method are otherwise larger for the simulations using the SRO EOS when using FLASH. This can be linked to a difference in the gravity and hydrodynamic treatment between the two codes, leading to a different PNS evolution (see Fig.~\ref{fig:1d_full_hydro}). 

This section presented the hydrodynamic and $\nu_x$ quantities for our different simulations depending on the code, EOS, pair processes treatment, and nucleon-nucleon bremsstrahlung formalism used. We show the close agreement between the full and approximate transport using the modified absorption opacity $\kappa^*_a$ for pair processes in Fig.~\ref{fig:lumi_factor}, as well as in simulations using GR1D. Simulations using the modified absorption opacity  $\kappa^*_a$ for the pair processes do not only reproduce the $\nu_x$ luminosity and mean energy of simulations using the full transport, they also show a great agreement in the radial neutrino distribution and luminosity, density, and temperature, for a considerably smaller computational cost.  Using the modified absorption opacity $\kappa^*_a$ in FLASH shows results in agreement with the ones found in simulations using GR1D, confirming the validity of our new approximation. However, the differences arising from a different transport or nucleon-nucleon bremsstrahlung formalism are small compared to differences arising from using a different EOS. We performed a set of two-dimensional simulations, which all lead to an explosion, to further study the impact of using a different EOS or $\nu_x$ transport and interactions.

\subsection{Neutrino effects in two dimensional simulations}\label{Sec:nu_effect_2d}

In this section, we will focus on the effect of the use of a different nucleon-nucleon bremsstrahlung formalism, OPE vs. T-matrix, and the effects linked to the transport method for the treatment of pair process interactions, the reference emissivity vs. the modified absorption opacity $\kappa^*_a$  in our two-dimensional simulations. We consider these effects assuming two different nuclear EOS.
\begin{figure}[h]
    \centering
    \includegraphics[width=\linewidth]{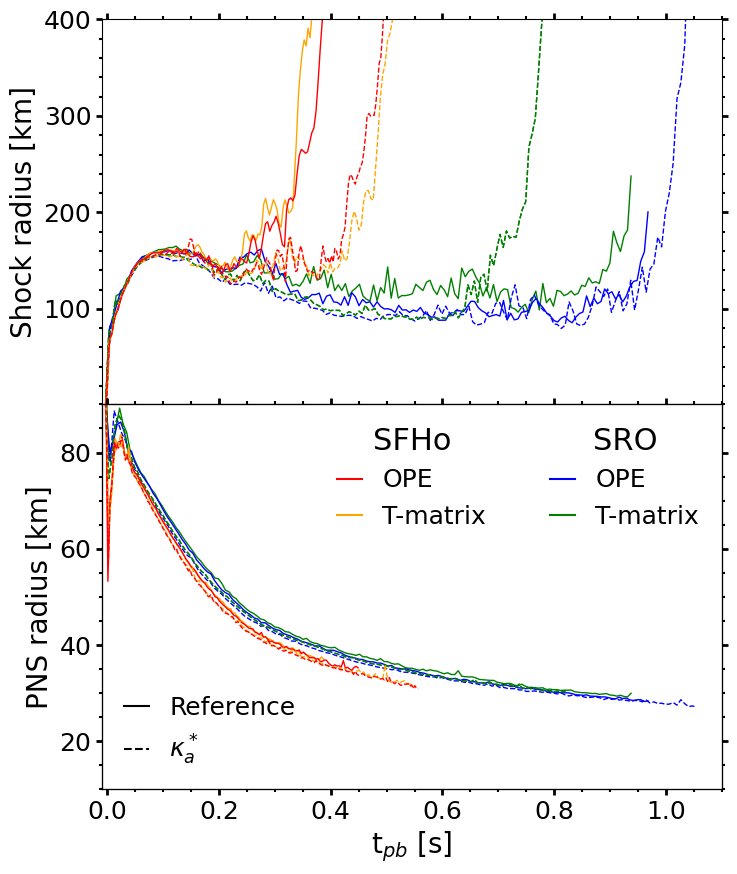}
    \caption{Shock evolution for all 2D simulations. The shock evolution and explosion time are extremely dependent on the EOS, with a difference of up to $\sim$500\,ms between explosion times.}
    \label{fig:full_2D_explo}
\end{figure}
In order to compare each effect, we performed a set of two-dimensional simulations, each of which ran until the time of the explosion. In Fig.~\ref{fig:full_2D_explo}, we show the shock radius evolution for our 2D simulations. The bounce time is different for the simulations using SRO and SFHo (319\,ms and 300\,ms, respectively); therefore, the time is considered time post-bounce using the respective bounce time.\\
Looking at the shock radius evolution, after the first $\sim$200\,ms, two families appear depending on the EOS used. The simulations using the SFHo EOS (red and orange lines) do explode near (or soon after) the time when the Si/O  shell interface accretes through the shock, between $\sim$350\,ms  and $\sim$500\,ms post-bounce. The simulations using the SRO EOS show a short shock revival after the first plateau as the Si/O shell is accreted, but it is insufficient to trigger the explosion, which will occur from $\sim$700\,ms post-bounce. The different evolutions of the simulations can be linked to the compactness of the PNS, with simulations using the SFHo EOS having a faster contraction, a denser core and a lower density in the gain region. \\

As the evolution of the explosion shows strong differences linked to the EOS, it is important to separate the simulations by EOS to observe the impact of neutrino transport and nucleon-nucleon bremsstrahlung formalism \cite{nagakura_systematic_2020,burgio_g_f_neutron_2021}. The different explosion times could lead to different quantitative conclusions on the impact of the neutrino effect in these simulations with different EOS, from the early explosion in simulations based on the SFHo EOS to the late explosions for simulations based on the SRO EOS. We, therefore, begin with a study of the impact of the neutrino effects in the simulations based on the SFHo EOS and then study the impact of these effects in simulations based on the SRO EOS. While, here, we mainly consider the EOS as being the driving factor for the hydrodynamic evolution differences, it is important to point out that these kinds of differences can also be linked to stochastic effects, as pointed out in \cite{oconnor_two-dimensional_2018,couch_revival_2013}.

\subsubsection{SFHo}\label{subsec:SFHo}

\begin{figure}
    \centering
    \includegraphics[width=\linewidth]{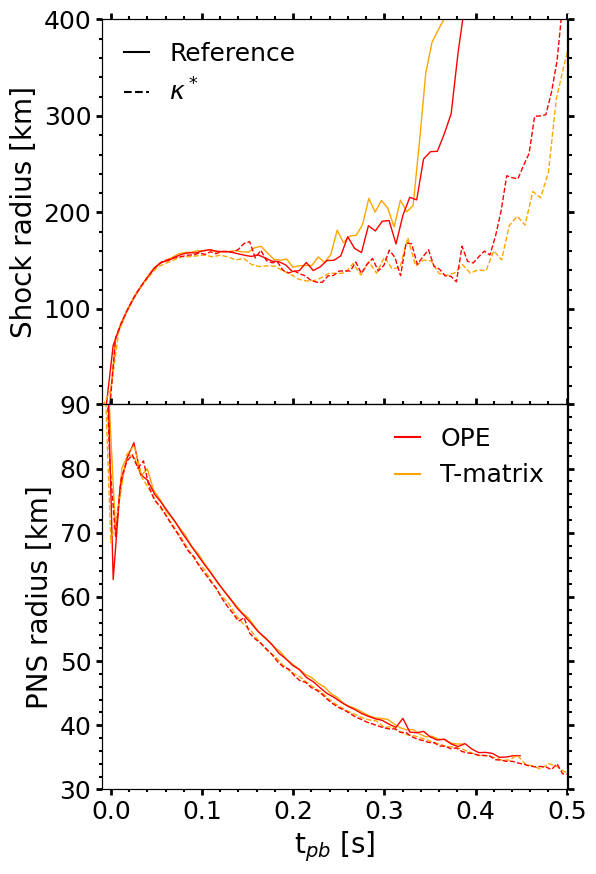}
    \caption{Shock radius (top) and PNS radius (bottom) evolution for the 2D simulations using the SFHo EOS. In red are the simulations using the OPE formalism for the nucleon-nucleon bremsstrahlung; in orange are the simulations using the T-matrix formalism. The solid lines represent the simulations using the old approximation for the pair processes, and the dashed lines represent the new approximation using a modified opacity $\kappa^*_a$. }
    \label{fig:hydro_SFHo_2D}
\end{figure}

In this subsection, we present the results of our 2D simulations based on the SFHo EOS. We begin by presenting the hydrodynamic results in Fig.~\ref{fig:hydro_SFHo_2D}. In this figure, we show the shock radius evolution with time in the top panel and the PNS radius evolution in the bottom panel. Early on, all trends are consistent with the observations from our 1D simulations. The red curves represent simulations using the OPE formalism for the nucleon-nucleon bremsstrahlung, and the orange curves represent simulations using the T-matrix formalism. The solid lines are simulations using the reference absorption opacity, and the dashed lines represent simulations using our modified absorption opacity $\kappa^*_a$ for the neutrino pair processes transport. Focusing first on the shock radius evolution, all the simulations show a monotonic shock expansion, reaching a plateau of $\sim$150\,km around 60\,ms post-bounce. This plateau lasts for $\sim$100\,ms before showing a small recession followed by a rapid expansion around the time of Si/O shell interface accretion. These 2D simulations show that the modified absorption opacity $\kappa_a^*$, which captures the $\nu_x$ transport more faithfully than the reference method, is systematically impacting the simulated non-linear evolution of the post-bounce core-collapse phase, including the time of the explosion. Due to this complex non-linearity, it is difficult to pin point the precise cause, nevertheless we present a thorough analysis of various aspects of these simulations in search for other systematic differences that might be cause by the $\nu_x$ transport approximations.

We show the PNS radius evolution on the bottom panel of Fig.~\ref{fig:hydro_SFHo_2D}. Around 20\,ms post-bounce, the PNS radius reaches a peak of $\sim$ 80\,km for all simulations. When the PNS begins to contract (at $\sim$50\,ms post-bounce), simulations using the modified absorption opacity $\kappa^*_a$ show a PNS radius consistently lower by a few percent. While the difference does not exceed $\sim$2\,km, it is consistent with a faster cooling of the PNS brought by a higher $\nu_x$ luminosity. \\

\begin{figure}
    \centering
    \includegraphics[width=\linewidth]{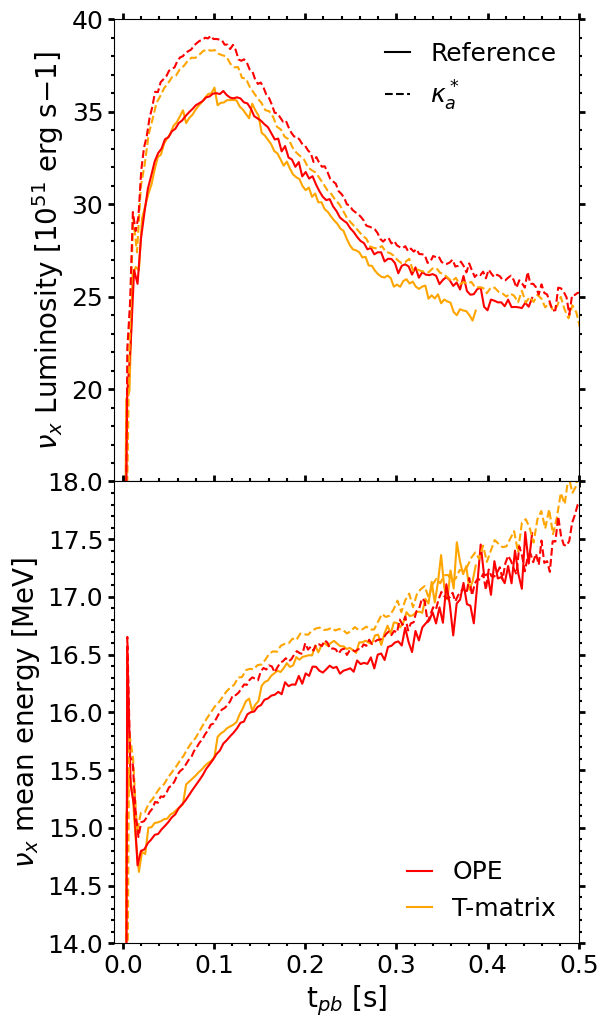}
    \caption{heavy-lepton neutrino luminosity (top) and mean energy (bottom) evolution for the 2D simulations using the SFHo EOS. The line type and colors are the same as for Fig.~\ref{fig:hydro_SFHo_2D}. }
    \label{fig:nu_SFHo_2D}
\end{figure}

In Fig.~\ref{fig:nu_SFHo_2D}, we present the heavy-lepton neutrino luminosities (top) and mean energies (bottom) for our 2D simulations using the SFHo EOS. The luminosities show a peak at $\sim$100\,ms post-bounce, reaching a maximum value of $\sim$ 35--40 $\times$ 10$^{51}$\,erg\,s$^{-1}$. This peak is followed by a linear decrease up until the time of the explosion. The use of the modified absorption opacity $\kappa^*_a$ increases the peak luminosity peak by $\sim$10\%, but this difference decreases after $\sim$200\,ms and becomes negligible. The impact of a different formalism for the nucleon-nucleon bremsstrahlung interaction is lower ($\sim$2\%) but stays constant throughout the simulation. These results are similar to the ones observed in Section.~\ref{sec:1D}.  The difference in luminosity is consistent with the previously observed PNS radius difference.\\

The $\nu_x$ mean energy shown in the bottom panel of  Fig.~\ref{fig:nu_SFHo_2D} increases linearly throughout the simulation while the PNS contracts. This is due to an increase in the temperature at the $\nu_x$ neutrinosphere, which leads to the production of higher energy neutrinos. The mean energy shows a short peak at bounce, reaching $\sim$16.5\,MeV, before dropping to a minimum value of $\sim$14.5\,MeV at $\sim$~20\,ms post-bounce, before increasing linearly up to $\sim$18\,MeV. Using a modified absorption opacity $\kappa^*_a$ increases the mean energy by $\sim$5\% steadily throughout the simulation. Using the T-matrix formalism for nucleon-nucleon bremsstrahlung shows little impact for the first 150\,ms, and then increases the luminosity by $\sim$5\% in a similar fashion to the transport method, once again consistent with 1D results. \\

 \begin{figure}
     \centering
     \includegraphics[width=\linewidth]{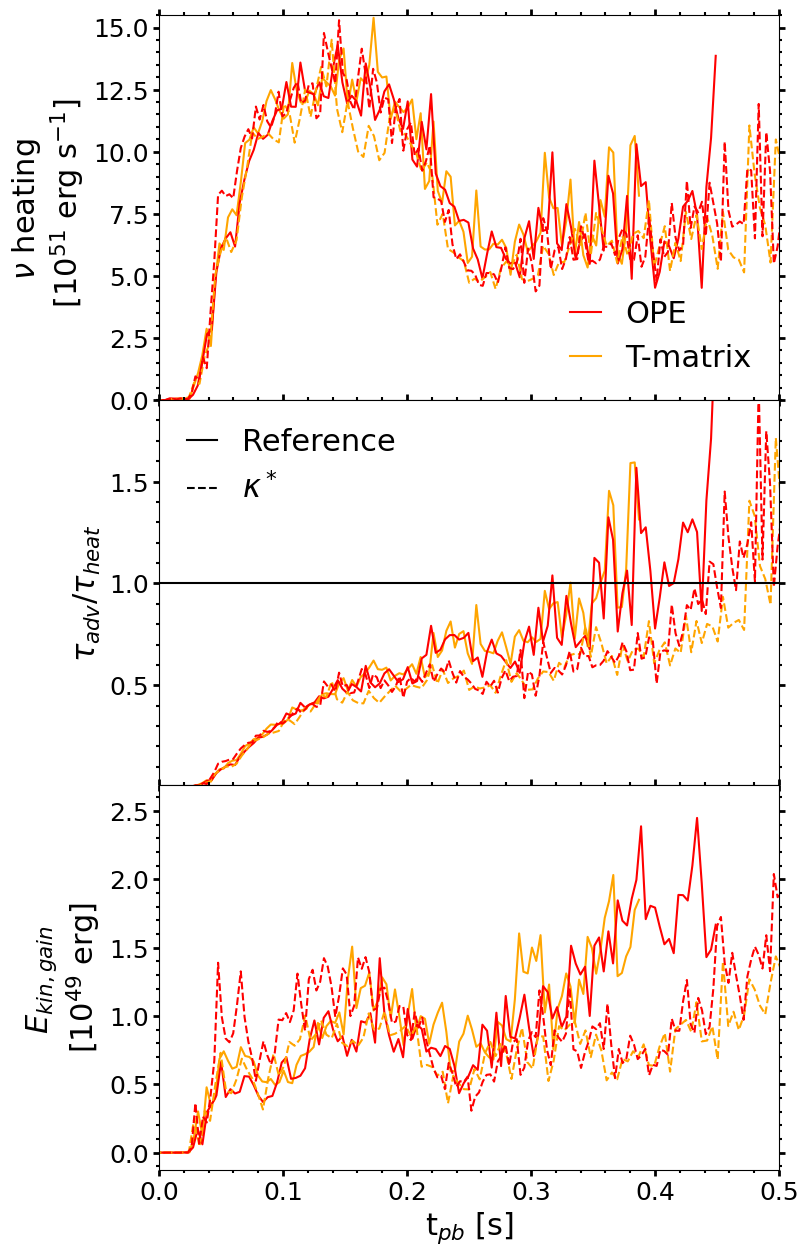}
     \caption{Net neutrino heating in the gain region (top panel),  $\tau_\mathrm{adv}/\tau_\mathrm{heat}$ (middle panel), and kinetic energy in the gain region (bottom panel) for simulations using the SFHo EOS. Line style and color are the same as the ones used in Fig.~\ref{fig:hydro_SFHo_2D} and presented in Tab.~\ref{tab:linestyle-color}. }
     \label{fig:SFHo_nu_heat}
 \end{figure}

As the transport of $\nu_e$ and $\bar{\nu}_e$ is the main mechanism of energy transport between the cooling PNS and the matter behind the shock, they are the main factors influencing the heating. The net neutrino heating is shown in Fig.~\ref{fig:SFHo_nu_heat} (top panel), along with the ratio of the advection and heating timescale (middle panel), and the kinetic energy in the gain region. The ratio of the advection and heating timescale is commonly used to represent the explosion trigger as a heating timescale that is smaller than the advection timescale allows the matter to expand. $\tau_\mathrm{adv}/\tau_\mathrm{heat}$ shows a difference between the simulations starting from $\sim$200\,ms post-bounce based on the pair processes transport method, with simulations using the modified opacity $\kappa_a^*$ having a lower $\tau_\mathrm{adv}/\tau_\mathrm{heat}$ than simulations using the reference transport. These differences in $\tau_\mathrm{adv}/\tau_\mathrm{heat}$ last up until the explosion triggers, with $\tau_\mathrm{adv}/\tau_\mathrm{heat}$ reaching 1 at $\sim$350\,ms post-bounce for simulations using the reference opacity, while for simulations using the modified absorption opacity $\kappa_a^*$, $\tau_\mathrm{adv}/\tau_\mathrm{heat}$ reaches 1 at $\sim$450\,ms post-bounce. The net neutrino heating shows little to no differences between the simulations, but the kinetic energy in the gain layer is higher for simulations using the modified absorption opacity $\kappa_a^*$ than for simulations using the reference transport through the first $\sim$200\,ms. This inverses after $\sim$250\,ms post-bounce, with simulations using the reference transport having a higher kinetic energy in the gain layer, following the same trend as $\tau_\mathrm{adv}/\tau_\mathrm{heat}$.

While $\nu_x$ neutrino luminosities and mean energies show differences depending on the transport method and nucleon-nucleon bremsstrahlung formalism, this difference not as straight-forward for electron neutrinos and anti-neutrinos. As we apply the pair processes only on heavy-lepton neutrinos (due to their low impact on electron neutrino and anti-neutrino quantities), the change in electron neutrino and anti-neutrino energy and luminosity is stemming purely from the hydrodynamic conditions at the neutrinosphere and the gain region. If plotted, the electron neutrino and anti-neutrino luminosities at 500\,km and their mean energies do not show significant differences, explaining the similar net heating in all simulations. Nevertheless, there could be systematic differences in the radial profile of the neutrino cooling and heating causing the observed trends between the various simulations. This is difficult to disentangle due to stochastic convection and turbulence occurring within the 2D simulations.\\

\begin{figure}
    \centering
    \includegraphics[width=\linewidth]{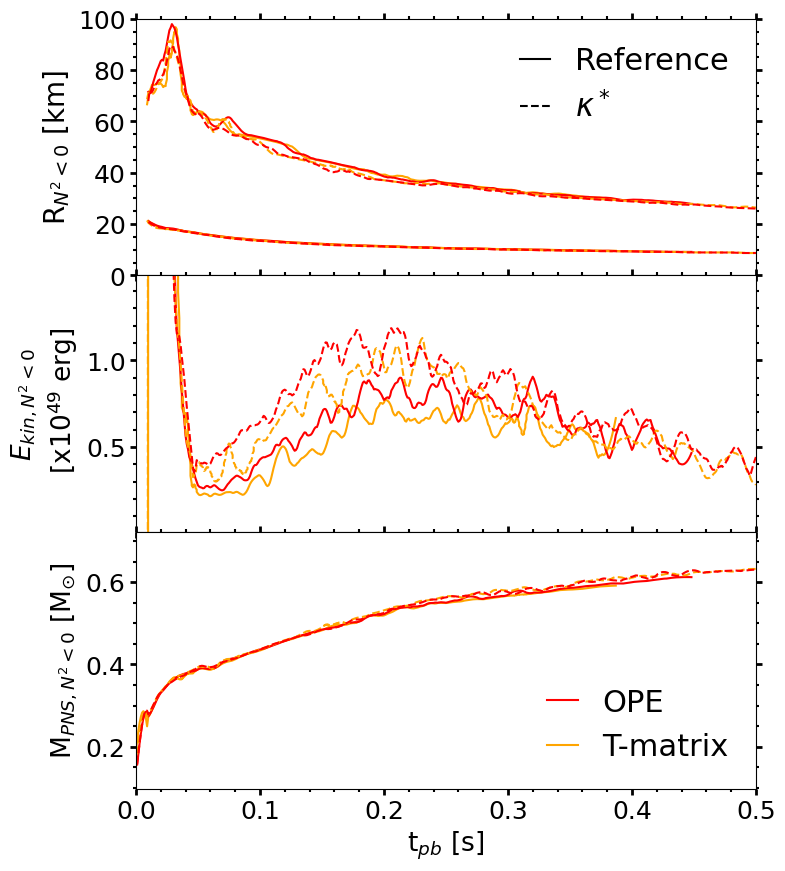}
    \caption{Radius (top panel), kinetic energy (middle panel), and mass (bottom panel) of the convectively unstable zone around the PNS manifested by a negative Brunt-V\"ais\"al\"a frequency. The top panel represents the radius of the bottom and top of the convectively stable layer. The colors and line styles are the same as the ones used in previous figures in this section and presented in Tab.~\ref{tab:linestyle-color}. }
    \label{fig:conv-PNS}
\end{figure}

We show in Fig.~\ref{fig:conv-PNS}  the radius, kinetic energy, and mass of the PNS convection zone. We define this convection zone as the entire convectively unstable layer starting in the PNS, but not constrained by the density limit of the PNS. This convectively unstable layer is defined as the layer in which the averaged Brunt-V\"ais\"al\"a frequency is consistently negative, following the Schwarzschild criterion \cite{aerts_asteroseismology_2010,mueller_new_2013}. Initially, this zone is quite large due to the prompt convection, but soon ($\sim50$\,ms) after bounce the entire convectively unstable zone is located inside the $\rho > 1\times 10^{11}$\,g\,cm$^{-3}$ limit, confirming it as the convection layer of the PNS. The radius of the convection layer is similar for all the simulations, with simulations using the reference transport approximation having a slightly larger convection zone for the first 200\,ms post-bounce. The main difference appears in the kinetic energy, with simulations using the modified absorption opacity $\kappa^*_a$ formalism having a kinetic energy higher by up to $\sim$30\,\% from 50\,ms to $\sim$300\,ms post-bounce. The simulation using the modified absorption opacity $\kappa^*_a$ transport coupled with the OPE formalism for nucleon-nucleon bremsstrahlung shows the highest kinetic energy by while the simulation using the reference transport method and the T-matrix formalism for nucleon-nucleon bremsstrahlung shows the lowest kinetic energy. This difference impacts the hydrodynamic quantities in the PNS convection zone, which is around the electron neutrino and anti-neutrino neutrinospheres, and therefore may impact the cooling, heating, and the emergent energy and luminosity of these neutrinos \cite{keil_ledoux_1996,dessart_multidimensional_2006,buras_two-dimensional_2006}. A higher temperature and density at the neutrinosphere explains the fact that a faster contraction of the PNS tends to lead to earlier explosions. While this faster contraction could be counted as an advantage toward the explosion, in our simulations, the kinetic energy in the gain region at the time of explosion ($\sim $200\,ms) is smaller for simulations using the modified absorption opacity $\kappa_a^*$ and the shock revival occurs later (see Fig.~\ref{fig:SFHo_nu_heat}).

\subsubsection{SRO}\label{subsec:SRO}

In this section, we will present the results from our simulations based on the SRO EOS.  
As in Sec.~\ref{subsec:SFHo}, we used FLASH to perform a set of 2D simulations, using different transport methods and nucleon-nucleon bremsstrahlung formalisms.  We begin by presenting the hydrodynamic quantities such as the shock and PNS radius, followed by the neutrino luminosities and mean energies, and finally a closer study of the PNS convection layer. 

\begin{figure}
    \centering
    \includegraphics[width=\linewidth]{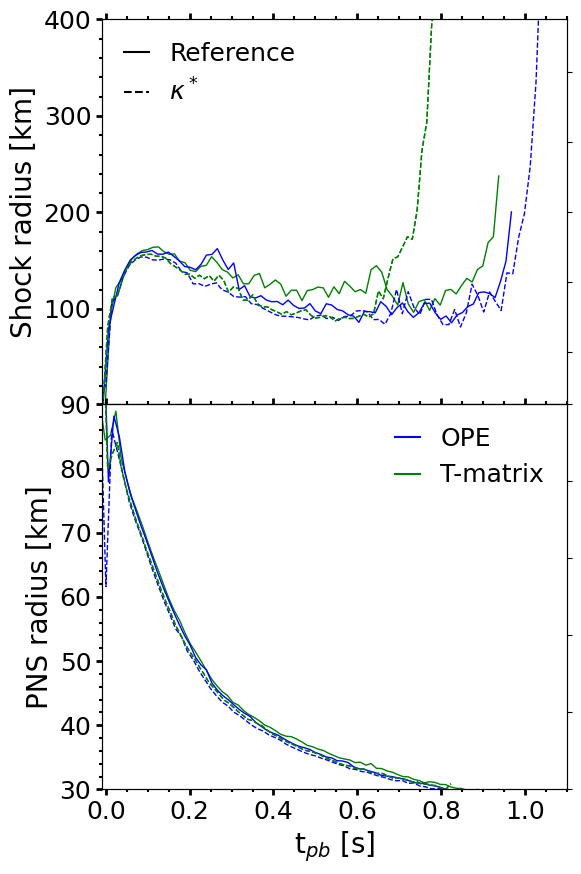}
    \caption{Shock radius (top) and PNS radius (bottom) evolution for the 2D simulations using the SRO EOS. In blue, the simulations use the OPE formalism for the nucleon-nucleon bremsstrahlung, in green the simulations use the T-matrix formalism. The solid lines represent the simulations using the old approximation for the pair processes, and the dashed lines represent the new approximation using a modified opacity $\kappa^*_a$.  }
    \label{fig:hydro_SRO_2D}
\end{figure}

In Fig.~\ref{fig:hydro_SRO_2D}, we show the shock radius evolution (top panel) and the PNS radius evolution (bottom panel). As in Sec.~\ref{sec:1D}, simulations plotted in blue use an OPE formalism for the nucleon-nucleon bremsstrahlung interaction, while the simulations plotted in green use a T-matrix formalism for this interaction. The simulations plotted in solid lines use the reference opacity for the simplified transport, while those plotted in dashed lines use our modified opacity $\kappa^*_a$.  \\
The shock radius expands for the first $\sim$50ms before plateauing and then splitting into two families, with the simulations using the reference opacity having a slightly larger radius than the ones using the modified opacity $\kappa^*_a$. The shock radius then presents a short expansion at $\sim$250ms post-bounce when the Si/O shell is accreted. This difference still does not trigger the explosion in any of the cases and the shock recedes. It then stalls at $\sim$100\,km before expanding at different times depending on the simulation. The explosion time does not seem to depend on the transport method, but simulations using the T-matrix formalism seem to explode earlier than simulations using the OPE formalism. These differences, especially since they are occurring at late times, are of small amplitude and would likely be unnoticeable when compared to the addition of stochastic effect stemming from initial progenitor perturbations or post-bounce turbulence \cite{oconnor_two-dimensional_2018,betranhandy_neutrino_2022-1}. \\

The PNS radius evolution is shown in the bottom panel of Fig.~\ref{fig:hydro_SRO_2D}. The PNS radius peaks at $\sim$88\,km at 25ms post-bounce, with simulations using the reference opacity having a larger PNS radius by $\sim$2\,km. This difference stays throughout the simulation as the higher neutrino luminosity allows the PNS to cool and contract faster, similarly as for the 1D simulations presented in Sec.~\ref{subsec:SFHo}.

\begin{figure}
    \centering
    \includegraphics[width=\linewidth]{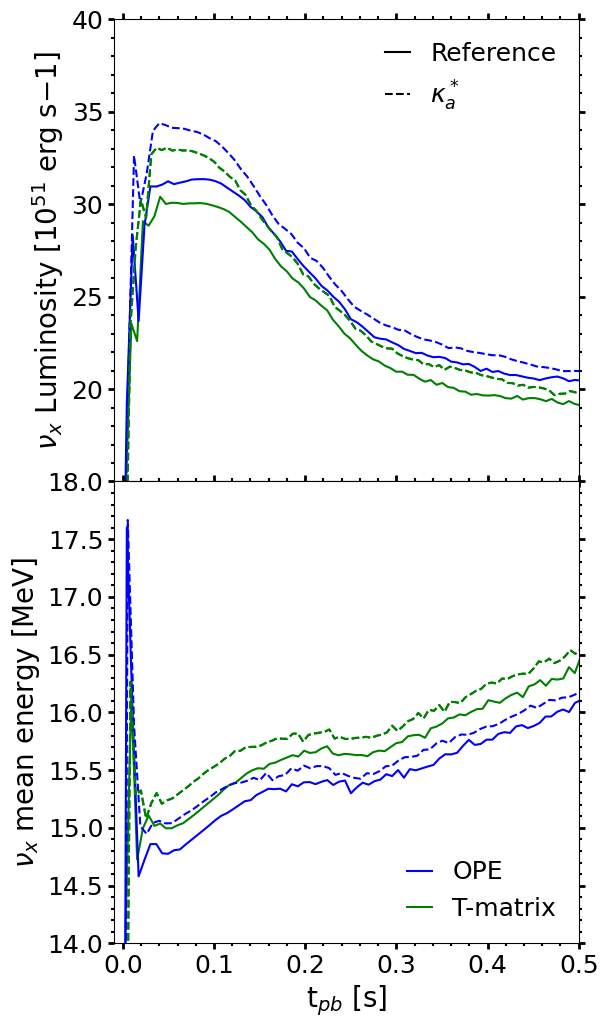}
    \caption{heavy-lepton neutrino luminosity (top) and mean energy (bottom) evolution for the 2D simulations using the SRO EOS. The line type and colors are the same as for Fig.~\ref{fig:hydro_SRO_2D} and presented in Tab.~\ref{tab:linestyle-color}. }
    \label{fig:nu_SRO_2D}
\end{figure}

In Fig. \ref{fig:nu_SRO_2D}, we present the $\nu_x$ luminosity (top panel) and mean energy (bottom panel) evolution with time. Similar to the simulation using the SFHo EOS presented in Sec.~\ref{subsec:SFHo}, simulations using the modified absorption opacity $\kappa^*_a$ show higher luminosity throughout the simulation. Comparing the same nucleon-nucleon bremsstrahlung formalism, the use of the modified absorption opacity $\kappa^*_a$ increases the luminosity through the simulation by $\sim$10\%. For differences linked to the bremsstrahlung formalism, simulations using the OPE formalism have a systematically higher luminosity, with a difference of up to $\sim$ 5\% at 50ms post bounce. This difference is lower than the one brought by the use of a different opacity, but compounding the two differences makes a maximum difference of $\sim$15\% in the luminosities. \\

A similar trend is shown in the mean energy (bottom panel), with simulations using the modified absorption opacity $\kappa^*_a$ having a higher mean energy for the heavy-lepton neutrinos higher by $\sim$3\% in the first 300ms post-bounce when comparing simulations using the same formalism for the nucleon-nucleon bremsstrahlung. The difference linked to the nucleon-nucleon bremsstrahlung formalism while using the same opacity brings a difference of $\sim$3\% in the mean energy, with simulations using the T-matrix formalism showing a higher mean energy. This makes the mean energy in the simulation using both the T-matrix formalism and the modified absorption opacity $\kappa^*_a$ higher by $\sim$6\% when compared to simulations using the OPE formalism and the reference opacity for pair processes. The difference between formalisms is more pronounced in simulations using the SRO EOS than in simulations using the SFHo EOS, but this is largely due to the difference in PNS structure, leading to different neutrinosphere radii. \\

 \begin{figure}
     \centering
     \includegraphics[width=\linewidth]{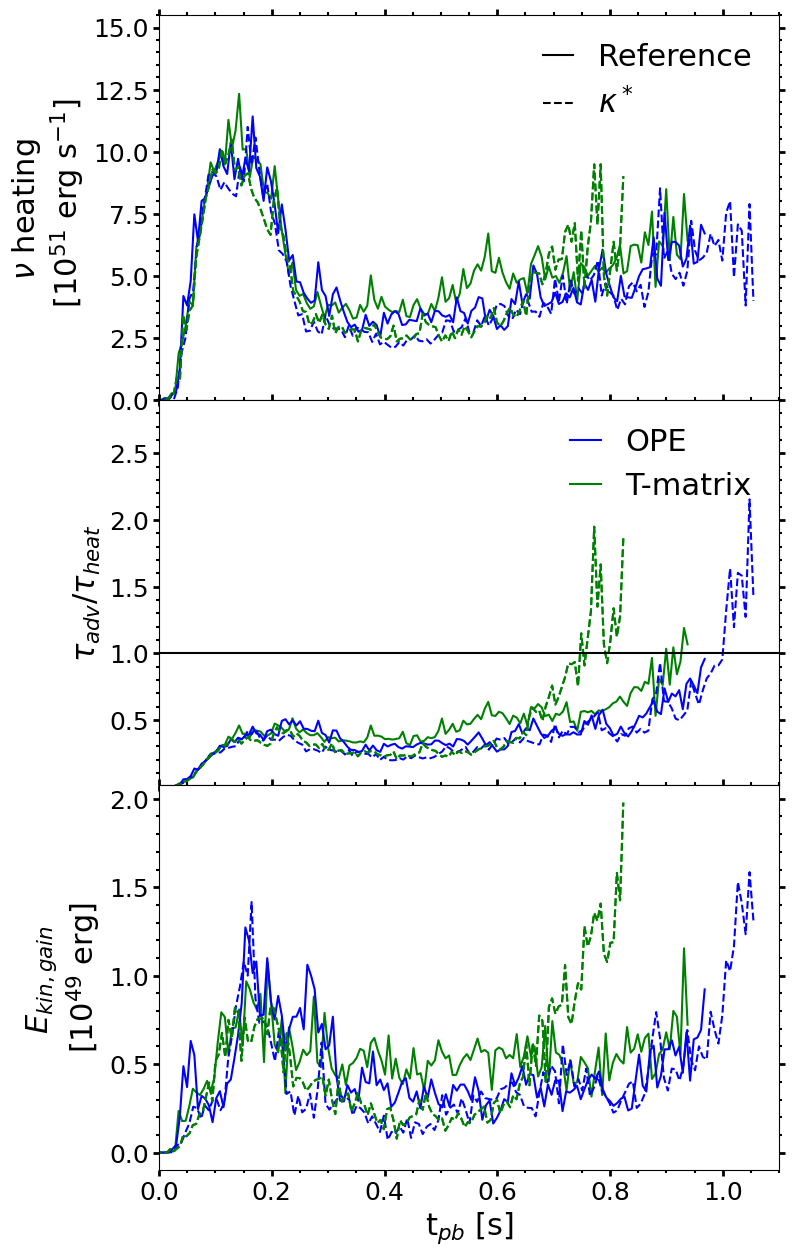}
     \caption{Net neutrino heating in the gain region (top panel),  $\tau_\mathrm{adv}/\tau_\mathrm{heat}$ (middle panel), and kinetic energy in the gain region (bottom panel) for simulations using the SRO EOS. Line style and color are the same as the ones used in Fig.~\ref{fig:hydro_SRO_2D}.}
     \label{fig:SRO_nu_heat}
 \end{figure}

These differences explain the different PNS radius evolution depending on the opacity and nucleon-nucleon bremsstrahlung formalism, but to see if it could impact the explosion time, we show in Fig.~\ref{fig:SRO_nu_heat} the neutrino heating (top panel), advection and heating timescale ratio (middle panel), and kinetic energy in the gain radius (bottom panel). The proxy $\tau_\mathrm{adv}/\tau_\mathrm{heat}$, representing an explosion when reaching 1, shows differences up to 400\,ms between simulations. When comparing $\tau_\mathrm{adv}/\tau_\mathrm{heat}$ evolution to the heating evolution, we see a clear impact of the heating in the first $\sim$500\,ms post-bounce, with simulations using the reference opacity having a larger heating. This difference in the neutrino heating then disappears. This shows that the different explosion times (all after 700\,ms post-bounce) cannot be solely linked to the difference in neutrino heating. This difference is then likely linked to differences in the turbulence and convection in the gain layer. The bottom panel of Fig.~\ref{fig:SRO_nu_heat} represents the kinetic energy in the gain layer, which follows a similar trend as the heating for the first 500\,ms post bounce, with the kinetic energy only increasing with the explosion. We note that the peak heating is $\sim$20\,\% lower for these simulations using the SRO EOS when compared with the SFHo simulations in \ref{subsec:SFHo}. This, in part, explains the lack of an earlier explosion.

\begin{figure}
    \centering
    \includegraphics[width=\linewidth]{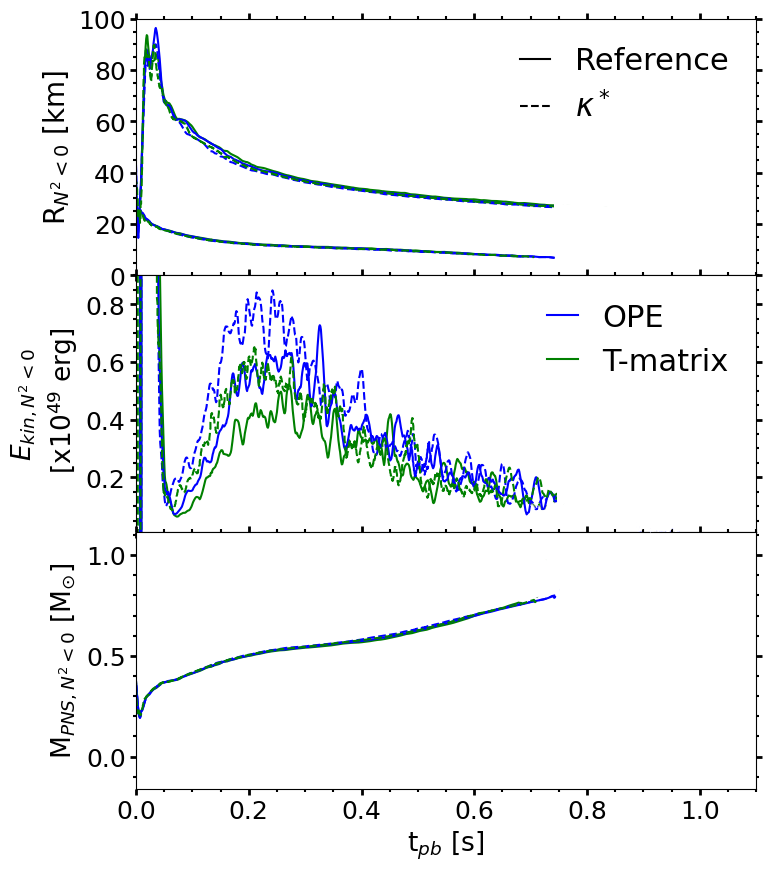}
    \caption{Hydrodynamic properties for the convectively unstable layer starting in PNS,  calculated as the presence of a negative Brunt-V\"ais\"al\"a frequency. The top panel represents the extent of the convective layer, the middle panel represents the kinetic energy in that layer, and the bottom panel represents the mass of the convective layer. The colors and line styles are the same as the one used in Fig.~\ref{fig:hydro_SRO_2D}. The scheme used to isolate the PNS convection layer is unreliable after $\sim$750\,ms post-bounce, and therefore we do not display the convection layer hydrodynamic properties after this point.}
    \label{fig:conv-PNS-SRO}
\end{figure}

In Fig.~\ref{fig:conv-PNS-SRO}, we show the radius, kinetic energy, and mass of the convectively unstable layer starting in the PNS. As mentionned in Sec.~\ref{subsec:SFHo}, this convectively unstable layer limits are inside the PNS after the first $\sim$50\,ms, confirming it as the convection layer of the PNS. The mass of the convection layer shows little difference between simulations using different treatments, while the kinetic energy and radius show differences from $\sim$75\,ms post-bounce. For the radius of the PNS convection layer, the differences are only distinguishable at $\sim$175 ms. Simulations using the modified opacity $\kappa_a^*$ show a higher kinetic energy in the PNS convection layer, and a slightly smaller radius of the convection zone. Simulations using the modified absorption opacity $\kappa^*_a$ transport coupled with the OPE formalism for nucleon-nucleon bremsstrahlung show the highest kinetic energy by $\sim$20\% when compared with simulations using the reference transport method and the T-matrix formalism for nucleon-nucleon bremsstrahlung. This difference impacts the density and temperature in the PNS convection zone, which is around the electron neutrino and anti-neutrino neutrinospheres, and therefore impacts the energy and luminosity of these neutrinos, while not being enough to change the course of the explosion as the net heating shows little differences.

\section{Conclusion}\label{sec:Conclusion}

In this paper, we presented the results of various modifications made to heavy-lepton neutrino transport processes.  We use one- and two-dimensional simulations as well as different nuclear EOS to quantify the impacts of these modifications. In Sec.~\ref{sec:method}, we presented a new approximation for an effective opacity used to model the pair-production of heavy-lepton neutrinos. This allows a better inclusion of neutrino physics without increasing the computational time. We also extend the results of \citetalias{betranhandy_impact_2020} by integrating the T-matrix formalism kernel for the nucleon-nucleon bremsstrahlung pair process, allowing us to use it as an effective opacity. 

In Sec.~\ref{sec:1D}, we presented the results of our new approximation in GR1D and FLASH, showing the accuracy of this modified absorption opacity $\kappa_a^*$ when compared to the full transport method in GR1D. The results show that our new approximation does a significantly better job at reproducing the results of the computationally-expensive full treatment when compared to the our original reference method used in \cite{oconnor_open-source_2015} and \citetalias{betranhandy_impact_2020} without increasing the computational cost. This shows that our modified absorption opacity $\kappa_a^*$ is a good replacement for the full calculation of the pair production processes for multidimensional simulations, where the computational cost is high.

To test the impact of our modified opacity in different scenarios depending on the explosion time, we used two different EOS, namely the SFHo EOS from \cite{steiner_core-collapse_2013} and the SRO EOS from \cite{schneider_equation_2020}, which was used as a baseline in \cite{andersen_equation--state_2021}. In our one-dimensional simulations, the impact of the new opacity and the use of a T-matrix formalism for the nucleon-nucleon bremsstrahlung is similar to the impact of using a different EOS. Simulations using the SRO EOS show a slower evolution of the PNS and a larger shock radius, leading to a lower $\nu_x$ luminosity and mean energy.  While in one-dimensional simulations, differences stemming from the use of different EOS or neutrino pair processes are in the same ballpark, the difference in PNS evolution depending on the EOS used makes for a vastly different explosion time in two-dimensional simulations for this progenitor. 

In Sec.~\ref{Sec:nu_effect_2d}, we showed the impacts of using our modified opacity and a different formalism for the nucleon-nucleon bremsstrahlung interaction in multi-dimensional simulations while separating simulations depending on the EOS used.

Using our modified absorption opacity $\kappa^*_a$ in simulations using the SFHo EOS results in a higher $\nu_x$ luminosity and mean energy, in turn leading to faster cooling of the PNS, which increases the kinetic energy in the convective layer of the PNS for the first $\sim$200\,ms post-bounce. As the neutrinospheres are located within the first $\sim$25\,km of the PNS, simulations using the modified absorption opacity $\kappa^*_a$ show a higher energy and luminosity for all the neutrino species. However, this effect only lasts for the first $\sim$200\,ms of the simulations, and does not impact the neutrino heating in a significant manner. This, coupled with the kinetic energy in the gain region being smaller for simulations using the modified absorption opacity $\kappa^*_a$, leads to a later shock revival. The use of a T-matrix formalism for the nucleon-nucleon bremsstrahlung did not impact the explosion in any systematic manner but did decrease the kinetic energy in the PNS convection layer when compared to the use of an OPE formalism. 

For simulations using the SRO EOS, the use of a different formalism for the nucleon-nucleon bremsstrahlung interaction and the use of our modified opacity did not make significant changes to the final explosion time. We do note that the use of the modified opacity $\kappa^*_a$ with the SRO EOS gave results consistent with the changes we saw for the SFHo EOS in the first 200-300\,ms (i.e. an increase in the the kinetic energy in the PNS convection zone, an increase in the emergent neutrino quantities, and a faster receding shock) but lower overall neutrino heating results in delayed (after 700\,ms) explosions.  At this later explosion time, no clear impact of the variations is seen.

The use of a different EOS leads to different shock evolution, as the difference around the neutrinospheres in both temperature and density heavily impacts neutrino emission, and thus the heating. Simulations using the SFHo EOS have a stronger neutrino heating, and a more compact inner region, leading to an earlier explosion. 

These different simulations show that the impact of a modification in the treatment of pair processes in $\nu_x$ can be seen in the case of an early explosion. In delayed explosions, using a different treatment does not significantly impact the explosion in this progenitor. It would, however, be interesting to study the impact of these differences on both the neutrino signal and the gravitational waves, as the PNS convection layer is heavily impacted by the neutrino treatment in the first 200\,ms post-bounce.

\section{Acknowledgements}

We want to thank Alexis Reboul-Salze for his insightful comments on this paper, and Sean Couch for FLASH software development. This work is supported by the Swedish Research Council (Project No. 2020-00452). Simulations were performed using the Yamazaki cluster from the CRA division at the Albert Einstein Institute, the Sakura cluster provided by Max Planck Computing and Data Facility, and the Tetralith cluster provided by the Swedish National Infrastructure for Computing (NAISS) at NSC partially funded by the Swedish Research Council through grant agreement no. 2024/3-32. 

\appendix
\section{}\label{sec:appendix}

In this appendix we motivate and calculate, in more detail than the main text, the derivation of our two different effective absorption opacities. We begin by considering the neutrino radiation moment source terms for a neutrino--anti-neutrino production/annihilation process \cite{shibata_truncated_2011},
\begin{equation}
S^\alpha_{\nu,\mathrm{pair}}=\nu^3 \int B^\mathrm{pair}_\nu (u^\alpha + l^\alpha )\,,
\end{equation}
where $\nu$ is the neutrino energy, $u^\alpha$ is the the four-velocity of the fluid, $l^\alpha$ is a unit vector perpendicular to $u^\alpha$, and the collision term, $B^\mathrm{pair}_\alpha$ is given as,
\begin{equation}
B^\mathrm{pair}_\nu = \int \nu'^2 d\nu' d\Omega' [ (1-f_\nu) (1-\bar{f}_\nu) R^\mathrm{pro} - f_\nu\bar{f}_\nu R^\mathrm{ann}] \,.
\end{equation}

After expanding the interaction kernels in moments and performing the angular integrals,  we arrive at a fairly complex expression in terms of the radiation moments of the two neutrinos involved, see Eq. 4.19 of \cite{shibata_truncated_2011}, repeated here,
\begin{widetext}
\begin{eqnarray}
\nonumber
S_{\nu,\mathrm{pair}}^{\alpha} &=& \int \frac{d\nu'}{\nu'} [-\{(J_{\nu} - 4\pi\nu^{3})u^{\alpha} + H_{\nu}^{\alpha}\} (4\pi \nu'^{3} -\overline{J}_{\nu'})R_{0}^{\mathrm{pro}}(\nu,\nu')\\
\nonumber
&&-\frac{\overline{H}_{\nu'}^{\alpha}}{3}]\{(4\pi\nu^{3} - J_{\nu})R_{1}^{\mathrm{pro}}(\nu,\nu')+ J_{\nu}\,R_{1}^{\mathrm{ann}}(\nu,\nu')\}\\
\nonumber
&&+(h_{\gamma \sigma}\,H_{\nu}^{\gamma}\,\overline{H}_{\nu'}^{\sigma}\,u^{\alpha} +\tilde{L}_{(\nu),\beta}^{\alpha} \overline{H}_{\nu'}^{\beta})
[\,R_{1}^{\mathrm{pro}}(\nu,\nu') - R_{1}^{\mathrm{ann}}(\nu,\nu')]\\
&&
-(J_{\nu}\,u^{\alpha} + H_{\nu}^{\alpha})\overline{J}_{\nu'}
R_{0}^{\mathrm{ann}}(\nu,\nu')].
\label{eq:fullsource}
\end{eqnarray}

\end{widetext}

This expression is used for the `Full' simulations in this paper. Considering the simplification of this equation by first, only retaining the 0$^\mathrm{th}$ order term in the kernel expansion (i.e.$R_1^{\mathrm{pro/ann}} = 0$), and second ignoring the final-state neutrino blocking associated with the neutrino pair production term this results in a greatly simplified expression,

\begin{eqnarray}
\nonumber
S_{\nu,\mathrm{pair}}^{\alpha} &=& \int \frac{d\nu'}{\nu'} [(4\pi\nu^{3}u^{\alpha})4\pi \nu'^{3}R_{0}^{\mathrm{pro}}(\nu,\nu')\\
&&
-(J_{\nu}\,u^{\alpha} + H_{\nu}^{\alpha})\overline{J}_{\nu'}
R_{0}^{\mathrm{ann}}(\nu,\nu')]\,,
\end{eqnarray}
which we then collect into a standard emission/absorption form,
\begin{equation}
S_{\nu,\mathrm{pair}}^{\alpha} = (\eta_{\nu} - \kappa_{\nu,a} J_\nu) u^\alpha - \kappa_{\nu,a} H_\nu^\alpha\,,
\label{eq:salphapair}
\end{equation}
where
\begin{equation}
\eta_\nu = 4\pi \nu^3 \int \frac{d\nu'}{\nu'} 4 \pi \nu'^3 R_0^\mathrm{pro}(\nu,\nu') \,,
\label{eq:emissivity_moment}
\end{equation}
and
\begin{equation}
\kappa_{\nu,a} = \int \frac{d\nu'}{\nu'} \overline{J}_{\nu'} R_0^\mathrm{ann}(\nu,\nu')\,.
\label{eq:kappanua}
\end{equation}
As we can see, with the approximations we have made so far, $\eta_\nu$ can be calculated a priori without knowing the particular neutrino distributions, on the other hand, $\kappa_{\nu,a}$ involves the entire energy distribution function of the other neutrino partner via  $\overline{J}_{\nu'}$ term in the integral expression, as to be expected since a neutrino of energy $\nu$ can annihilate with an anti-neutrino of any energy $\nu'$.

However, we know that if the neutrino fields are in equilibrium with the matter then this absorption term $\kappa_{\nu,a}$ must be equal to $\eta_\nu / J_{\nu, \mathrm{eq}}$ via Eq.~\ref{eq:salphapair}. For this case of neutrino-pair annihilation, this equality is only true when the neutrino fields are in equilibrium. Furthermore, this is equivalent to replacing the $\overline{J}_{\nu'}$ in Eq.~\ref{eq:kappanua} with $\overline{J}_{\nu',\mathrm{eq}}$, the equilibrium neutrino distribution of the partner neutrino.
In this paper, taking $\kappa_{\nu,a} = \eta_\nu / J_{\nu, \mathrm{eq}}$ is our `Reference' effective absorption opacity. In \citetalias{betranhandy_impact_2020} and \cite{oconnor_open-source_2015} it was shown that this approximation does a good job in capturing the transport of the heavy-lepton neutrinos. However, assuming that the partner neutrino has a distribution equal to the equilibrium distribution overestimates the absorption opacity (potentially up to $\sim$50\%, see \citetalias{betranhandy_impact_2020} and Fig.~\ref{fig:lumi_factor}). In this work, we extend this effective absorption opacity by applying a reduction factor to the opacity given by the ratio of the instantaneous energy density of the partner neutrino field to the equilibrium energy density of the partner neutrino field.  This correction factor must be calculated at each time step, but is done explicitly and is a straightforward integral. Our modified effective absorption opacity, `$\kappa^*_{a}$' is given as,
\begin{equation}
    \kappa^*_{\nu,a} = \frac{\Sigma_i \overline{E}_i \Delta \epsilon_i }{\Sigma_i \overline{J}_{i,\mathrm{eq}} \Delta \epsilon_i} \times \kappa_{\nu,a}\,.
\end{equation}
While Eq.~\ref{eq:salphapair} and this expression for the effective opacity still remains an approximation, as we show in this paper, it does a much better job at matching the transport results of simulations using the `Full' expression in Eq.~\ref{eq:fullsource} than simply using $\kappa_{\nu,a}$.

\bibliographystyle{abbrv}
\bibliography{total_library}
\end{document}